\documentclass[debug]{rmaa}

\usepackage{paralist}

\usepackage{psfrag,color}

\usepackage[latin1]{inputenc}

 \title{ Proper motions of L1551~IRS~5 binary system using 7~mm VLA observations} 

\author{
  A. M. Villa\altaffilmark{1}, 
  M. A. Trinidad\altaffilmark{1,3},
  E. de la Fuente\altaffilmark{2,3}
  and T. Rodr\'{\i}guez-Esnard\altaffilmark{2}}

\altaffiltext{1}{Departamento de Astronom\'{\i}a,Universidad de Guanajuato, M\'exico.}

\altaffiltext{2}{Departamento de Fisica, Universidad de Guadalajara, M\'exico.}

\altaffiltext{3}{Corresponding author: trinidad@astro.ugto.mx, edfuente@gmail.com}

\shortauthor{Villa et al.}

\shorttitle{Proper Motions in L1551~IRS~5}

\fulladdresses{
\item A. M. Villa, M. A. Trinidad and T. Rodr\'{\i}guez: Departamento de Astronom\'{\i}a, 
 Universidad de Guanajuato, Apdo. Postal 144, Guanajuato, Gto. 36240, M\'exico.

\item E. de la Fuente: Guadalajara, Jalisco, M\'exico.}

\listofauthors{A. Villa, M.A. Trinidad, E. de la Fuente, \& T. Rodr\'{\i}guez}
\indexauthor{Villa, A.}
\indexauthor{Trinidad, M. A.}
\indexauthor{de la Fuente, E.}
\indexauthor{Rodr\'{\i}guez, T.}

\abstract{We analyzed high angular resolution observations of the Very Large
Array archive
at a wavelength of 7~mm of the L1551~IRS~5 binary system. Six sets of observations,
five with the A configuration and one with the B configuration, were used, covering
a time span of about 15 years. With these multi-epoch data, we estimated the
absolute and relative proper motions of the binary system, which are about
25.1~mas~yr$^{-1}$ ($\sim 16.7$~km~s$^{-1}$ considering a distance
of 140~pc) and 4.2~mas~yr$^{-1}$, respectively. Finally, based on the relative
proper motion, we estimated a total mass of the L1551~IRS~5 binary system of 
1.7~M$_{\odot}$ and an orbital period of 246 years.}

\resumen{Analizamos observaciones de alta resoluci\'on angular del archivo
del Arreglo Muy Grande (Very Large Array) 
a una longitud de onda de 7~mm del sistema binario L1551~IRS~5 .  
Se usar\'on seis juegos de observaciones, cinco con la configuraci\'on A 
y uno con la B, cubriendo un intervalo de tiempo de alrededor de 15 a\~nos.
Con estos datos multi-\'epoca estimamos los movimientos propios absolutos
y relativos del sistema binario, los cuales son alrededor de 25.1~mas~yr$^{-1}$ 
($\sim 16.7$~km~s$^{-1}$ considerando una distancia de 140~pc) y 4.2~mas~yr$^{-1}$, respectivamente. Finalmente, en base al movimiento propio relativo estimamos
una masa total del sistema binario L1551~IRS~5 de 1.7~M$_{\odot}$ y un periodo 
orbital of 246 a\~nos.}

\addkeyword{Astrometry}
\addkeyword{ISM: Circumstellar disks}
\addkeyword{Stars: individual(L1551~IRS~5)}

\begin{document}

\maketitle

\section{Introduction}
\label{sec:intro}

The formation of isolated low-mass stars, within molecular clouds, is quite known.
The accretion model proposed by \citet{Shu1987} has been successfully tested both
theoretically and observationally. However, our understanding of the formation of binary
and multiple stellar systems is still far from having a robust model, as it has for
isolated low-mass stars. This is a really serious problem, since the most of stars 
are formed in this way.

The determination of proper motions has proven to be a powerful tool to study various
physical processes in regions of star formation. For example, the proper motions of
Herbig-Haro objects and molecular outflows have helped to improve our understanding of
the earliest evolutionary stages of the formation of single stars \citep[e.g.][]{Frank2014}. 
In this regard,
it is known that protostars also show proper motions, which, for near stars
are on the order of a few milli-arcseconds per year. In particular, the determination
of absolute and relative proper motions toward binary star systems allows us to study
the kinematics of both components, which could help to understand the interaction
between the binary system and its environment and thus, try to understand its process 
of formation \citep[e.g.][]{Rodriguez2004}.

The dark cloud Lynds 1551 (L1551) is located at a distance of about 140~pc and
hosts a very active star-forming region \citep[e.g.][]{Ungerechts1987}. In L1551
is embedded the IRS5 source, which is the brightest of the cloud (40 L$_{\odot}$) and a
young stellar object (YSO) prototype. However, L1551~IRS~5 is detected as a binary
system at radio wavelengths \citep{Bieging1985}. In particular, using observations
at 7 mm, \citet{Rodriguez1998} showed that the continuum emission is from two
compact protoplanetary disks, which are separated about 0.3" ($\sim$ 40~UA at a distance
of 140~pc). Furthermore, it is also known that L1551~IRS~5 is the driving source of a
molecular outflow \citep{Snell1980}.

Proper motion studies have already been performed toward the L1551~IRS~5 binary system. 
For example, \citet{Rodriguez2003}, \citet{Lim2006} and \citet{Lim2016}
 estimated proper motions of this system using VLA observations at 2~cm and, 
$2 + 0.7$~cm, respectively. However, the emission of the circumstellar disks at 
a wavelength of 2~cm has an important free-free contribution and a minimal 
contribution of dust emission, which does not trace only the emission of the 
circumstellar disks of L1551~IRS~5, but also of extended thermal jets associated with
the disks.
Since the emission of circumstellar disk comes from dust grains, and this one is
more compact than the emission of the radio jets, the most appropriate
wavelength of the VLA to determine the continuum emission of the circumstellar
disks, and so to determine their proper motions, is 7~mm.

In this paper, we analyze six sets of VLA observations of the continuum emission
at 7~mm to determine the absolute and relative proper motions of the circumstellar
disks of the L1551~IRS~5 binary system covering a time span of 15 years. The relative
proper motions, which are calculated by the L1551~IRS~5 system are the most accurate 
up to date, which allows for a good estimate of the total mass of the system and its period.
This paper is organized as follows, In Section 2 we describe the VLA archive observations
and in Section 3 we present the observational results. Discussion and Summary are given in
Section 4 and 5, respectively.

\section{The VLA archive Data: Observations and Data Reduction}
\label{sec:obser}

All observations, which were used to estimate the proper motions of the L1551~IRS~5 
binary system, were taken from the VLA archive of the National Radio Astronomy Observatory
(NRAO)\footnote{NRAO is a facility of the National Science Foundation operated
under cooperative agreement by Associated Universities, Inc.}. In order to
separate spatially the two components of the L151~IRS~5 and to avoid the contribution
of free-free emission to the compact circumstellar disks of L1551~IRS~5, we selected
only observations carried out with the A and B configuration of the VLA
at a wavelength of 7~mm. In addition, to determine reliable positions for the two
circumstellar disks of L1515~IRS~5, data sets with sufficient sensitivity
were only selected (rms $\le 0.2$~mJy). Under these considerations, we found five data sets at A 
configuration and one at B configuration in the VLA archive, which cover a large 
time span of about 15 years, from 1997 March to 2012 November. All observations
were performed in both left and right circular polarization with a bandwidth of 100~MHz.
Information of the six data sets is given in Table \ref{Table1}.

All data sets were edited, calibrated and imaged using the software Common Astronomy
Software Applications (CASA) following standard techniques.
In order to obtain the best radio continuum maps, we performed cleaned images of the field
using several ROBUST parameters of the CASA task CLEAN.
The phase calibrator and synthesized beam size of the multiepoch continuum maps
are given in Table \ref{Table1}. On the one hand, the positions of both circumstellar
disks of the binary system L1515~IRS~5 were obtained with the task IMFIT of CASA, 
which applies a Gaussian fit to the image (see Table \ref{Table2}).

\section{Results and Discussion}
\label{results}

\begin{figure}[ht]
  \includegraphics[angle=0,width=1.0\columnwidth]{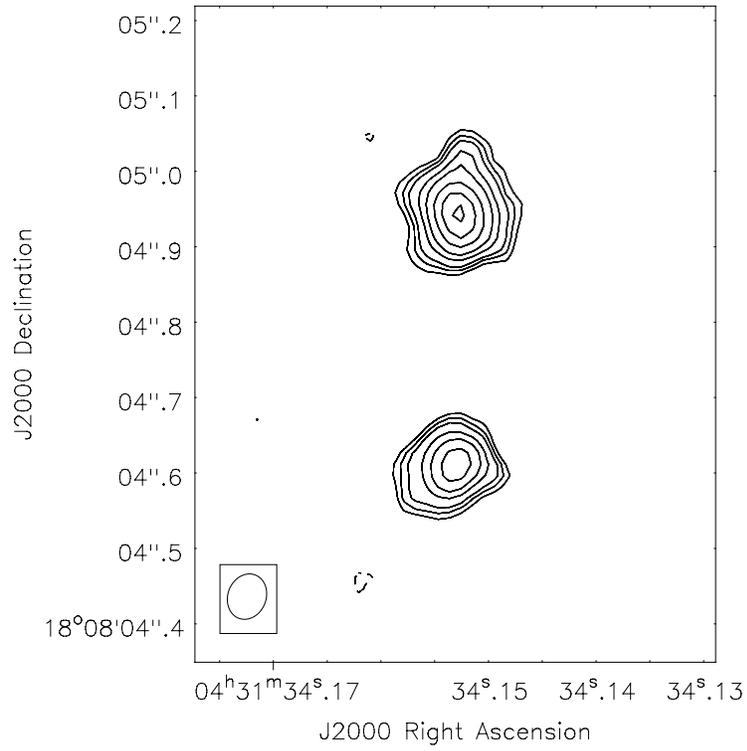}  
 \caption{Continuum contour map o the cicumstellar disk of the  L1551~IRS~5 
 binary system at 7~mm of the epoch 2003.73. Contours are -4, -3, 3, 4, 5, 7, 9, 12, 15 
 and 18 times 0.11~mJy~beam$^{-1}$, the rms noise of the map. The beam size is 
 $0.06 \times 0.05$ (bottom left corner).  }
  \label{Map1}
\end{figure}

\begin{figure}[ht]
  \includegraphics[angle=0,width=1.0\columnwidth]{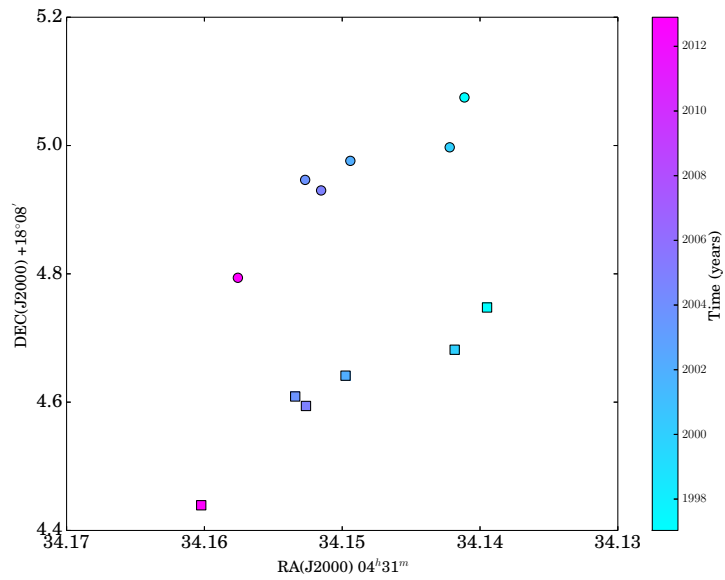}  
 \caption{Plot (right ascension, declination and time) of the temporal evolution
 of position of the compact circumstellar disks of the L1551~IRS~5 binary system
 over a time span of 15 years. 
 Circles and squares indicate the position of the north and south component, respectively.  }
  \label{Fig-3D}
\end{figure}

An example of 7~mm continuum map of the L1551~IRS~5 binary system (epoch 2003.73) 
is shown in Figure \ref{Map1}, where the spatial position of both compact 
circumstellar disks of L1551~IRS~5 is seen. On the other hand, Figure \ref{Fig-3D}
shows the temporal evolution of projected position of the two circumstellar disks 
(north and south components) over a time span of about 15 years. From this figure,
it is evident that the two components have clearly shifted their spatial position 
over time. Unfortunately, the six observation epochs found in the VLA archive
were not carried out using the same phase calibrator, three data sets used the 0431+206
phase calibrator (sets 1, 2 and 4), while the remaining three used the 0431+175 phase
calibrator (sets 2, 5 and 6; see Table \ref{Table1}). Given that the selected observations
used two different phase calibrators, it could not be advisable to use all data sets
simultaneously to determine the absolute proper motions of the two components of
L1551~IRS~5. However, all data sets can be used to calculate their relative proper 
motions.

\subsection{Absolute Proper Motions}
\label{APM}
\begin{figure*}[ht]
 \includegraphics[angle=0, width=0.55\linewidth,height=7.0cm]{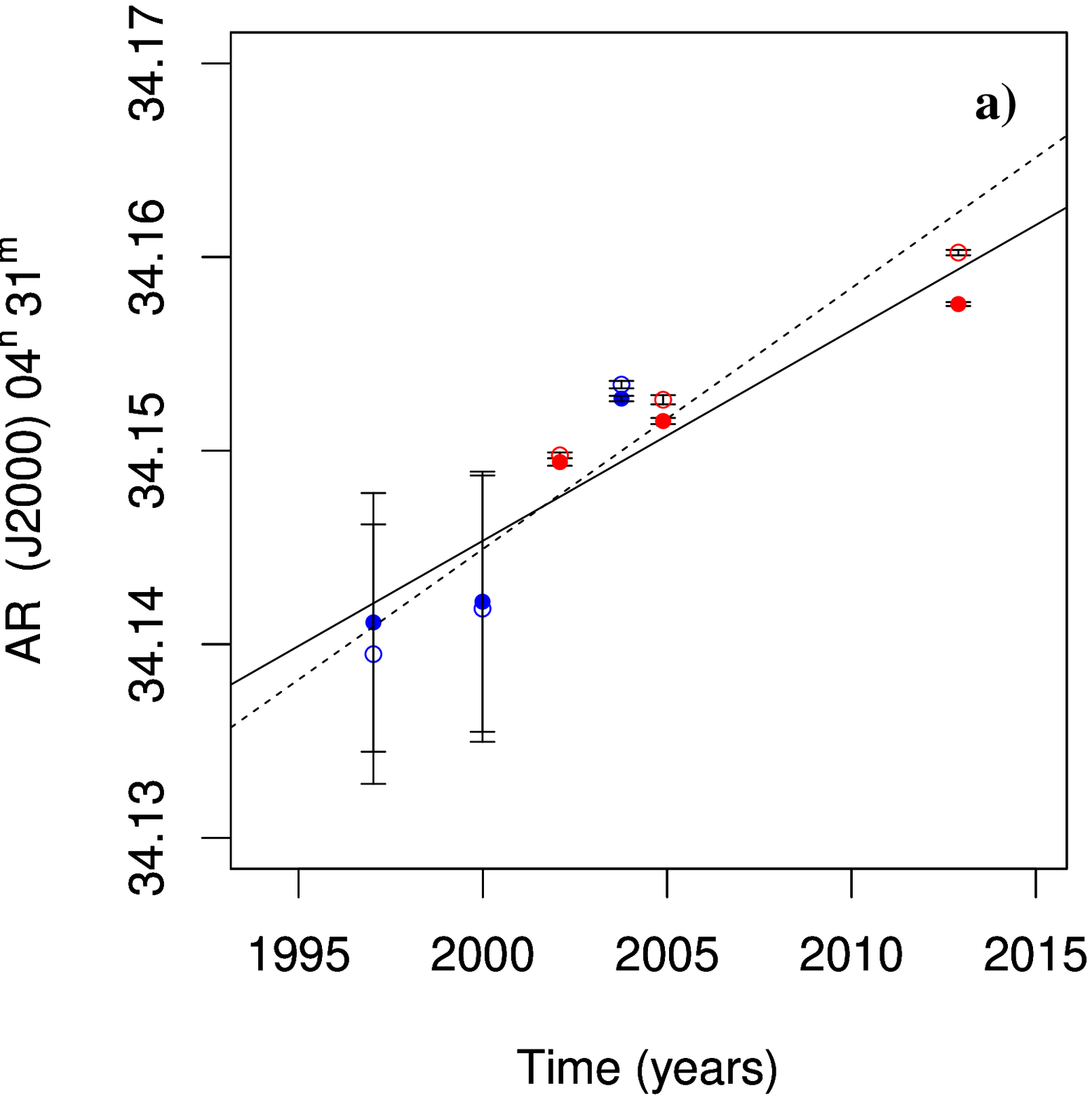} 
 \hfill
  \includegraphics[angle=0, width=0.55\linewidth,height=7.0cm]{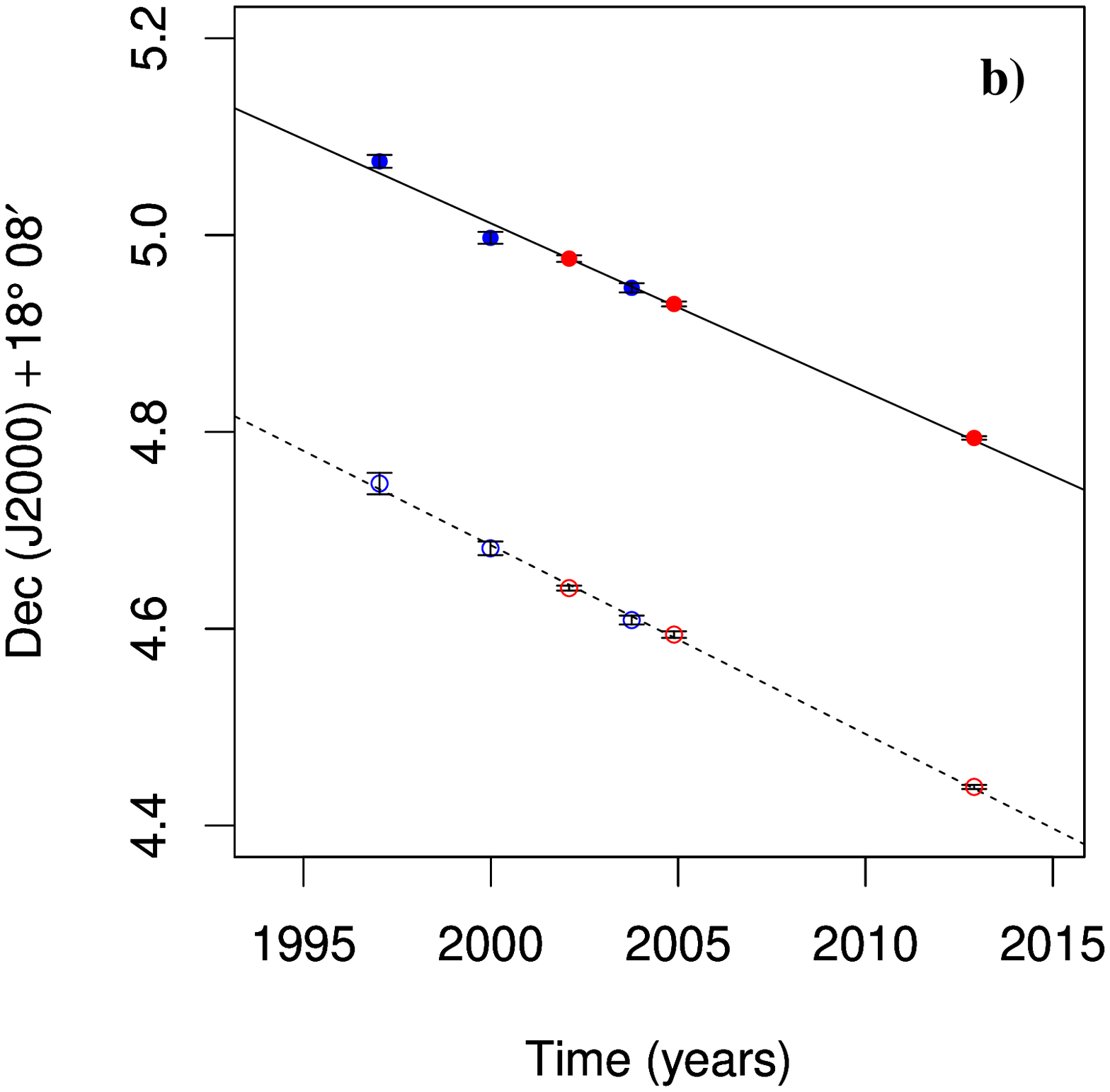}
   \caption{Absolute proper motions of north (full circles) and south (empty circles)
   components of the L1551~IRS~5 binary system using all six data sets,
   where blue circles are observations with the phase calibrator 0431+206 and red
   circles with the phase calibrator 0431+175 (error bars include both the positional
   uncertainty of the target source and the phase calibrator).
   Temporal evolution of the right ascension and declinations are shown in panel {\it a)} 
   and {\it b)}, respectively.  Solid and dashed lines represent the least-squares fits.
    }
  \label{Mov-Abs}
\end{figure*}

As we mentioned, two different phase calibrators were used in six data sets at
7~mm of L1551~IRS~5. Then, we are not able to calculate a reliable value for the
absolute proper motion of the L1551~IRS~5 system using all data sets. However,
given that the six data sets cover a long time span and both components of the
binary system are detected with a large signal-to-noise ratio, we can try to estimate
a very rough absolute proper motion for L1551~IRS~5 using all data sets by assuming
that the positional error of the phase calibrators (as observed by the VLA) is very
small. Under this assumption (and considering the positional uncertainty of the phase
calibrators), we made a linear least-squares fit to the measured
absolute positions of both components as function of time using the six epochs.
The results of these fits are shown in Figure \ref{Mov-Abs} (see also Table 
\ref{Table3}). The absolute proper motion of the L1551~IRS~5 binary system is 
on the order of  25.1~mas~yr$^{-1}$ ($\mu_{\alpha}=+17.4 \pm 4.0$~mas~yr$^{-1}$, 
$\mu_{\delta}=-18.2 \pm 0.6$~mas~yr$^{-1}$), which at a distance of 140~pc implies 
a velocity in the plane of the sky of $\sim 16.7$~km~s$^{-1}$. Very similar absolute
proper motions  were obtained  by \citet{Rodriguez2003}.

\begin{figure*}[ht]
 \includegraphics[angle=0, width=0.55\linewidth,height=7.0cm]{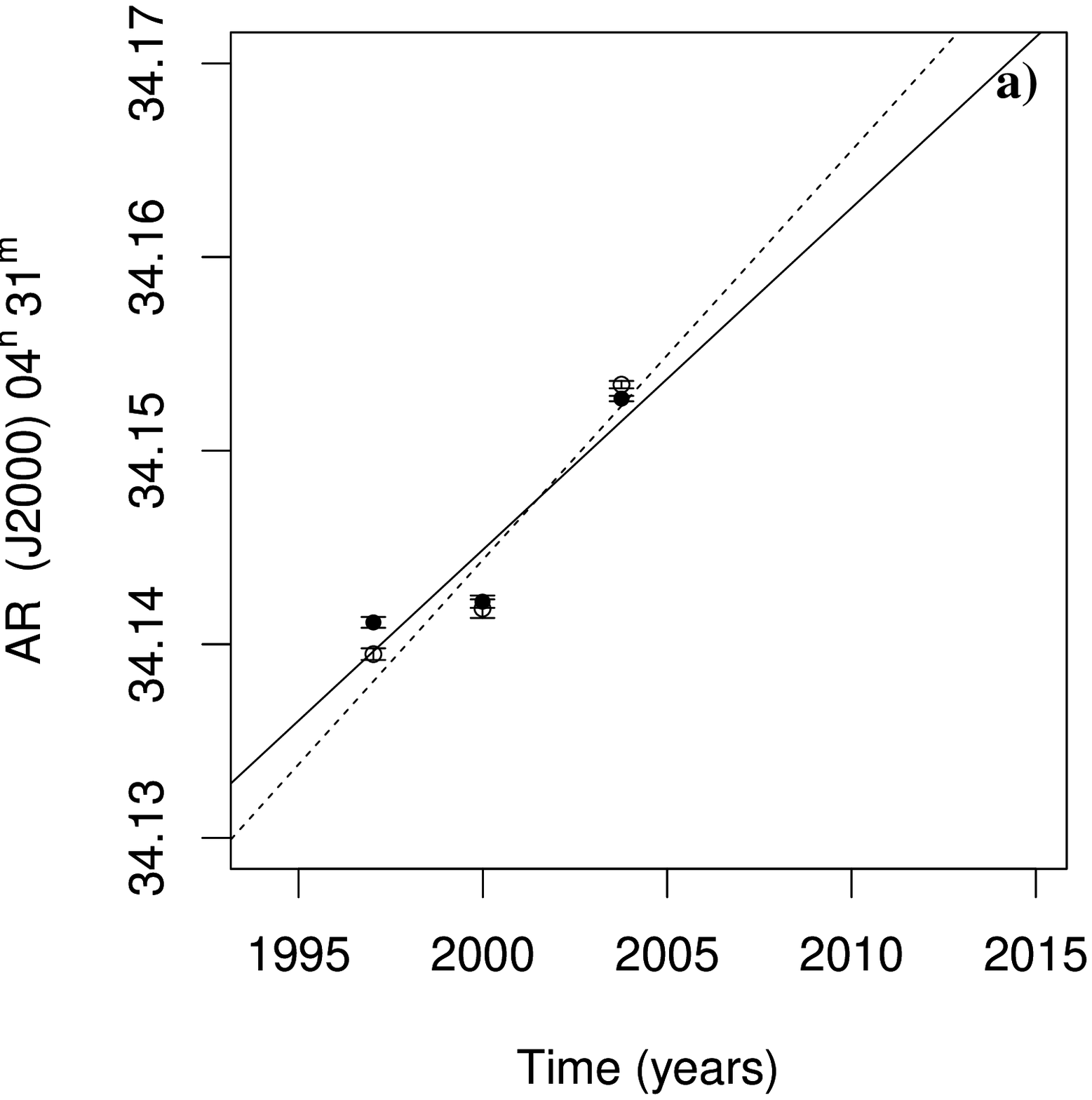} 
 \hfill
  \includegraphics[angle=0, width=0.55\linewidth,height=7.0cm]{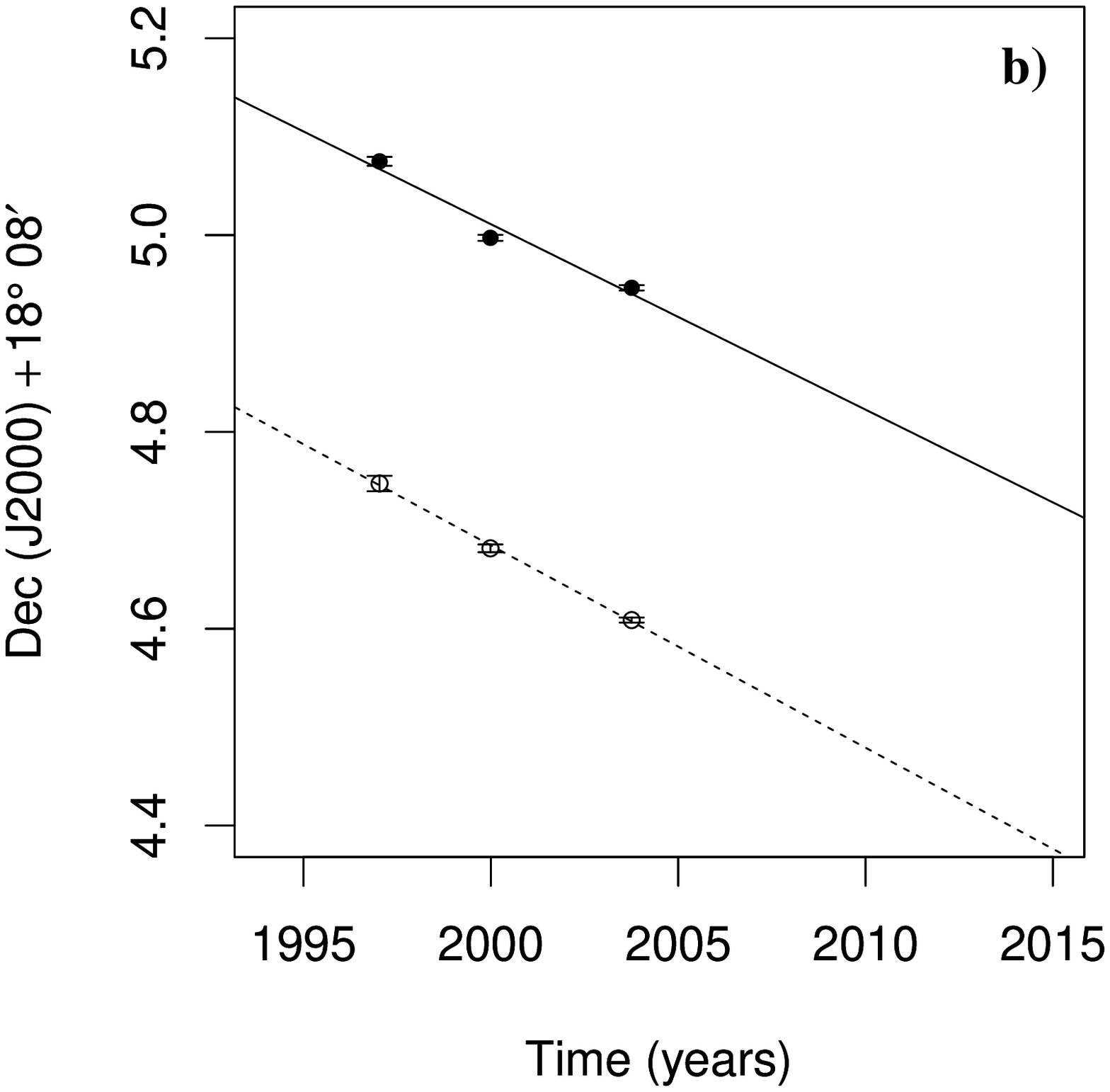}
   \caption{Absolute proper motions of north (full circles) and south (empty circles)
     components of the L1551 IRS 5 binary system for the subgroup 1 (data sets 1, 2 and 4).  
     Temporal evolution of the right ascension and declinations are shown in panel {\it a)} 
   and {\it b)}, respectively.  Solid and dashed lines represent the least-squares fits. }
  \label{Set-1}
\end{figure*}

\begin{figure*}[ht]
  \includegraphics[angle=0, width=0.55\linewidth,height=7.0cm]{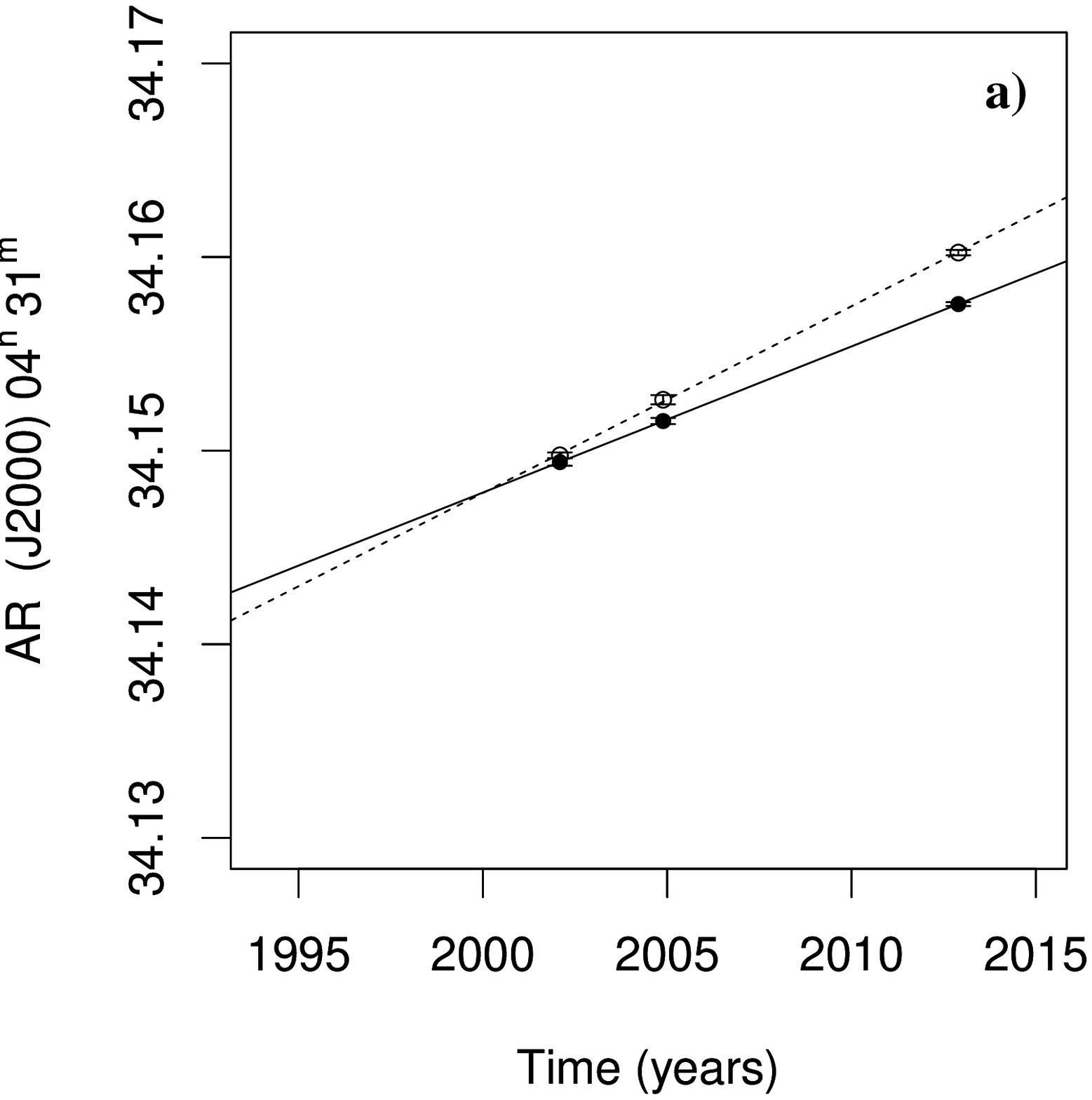}
 \hfill
  \includegraphics[angle=0, width=0.55\linewidth,height=7.0cm]{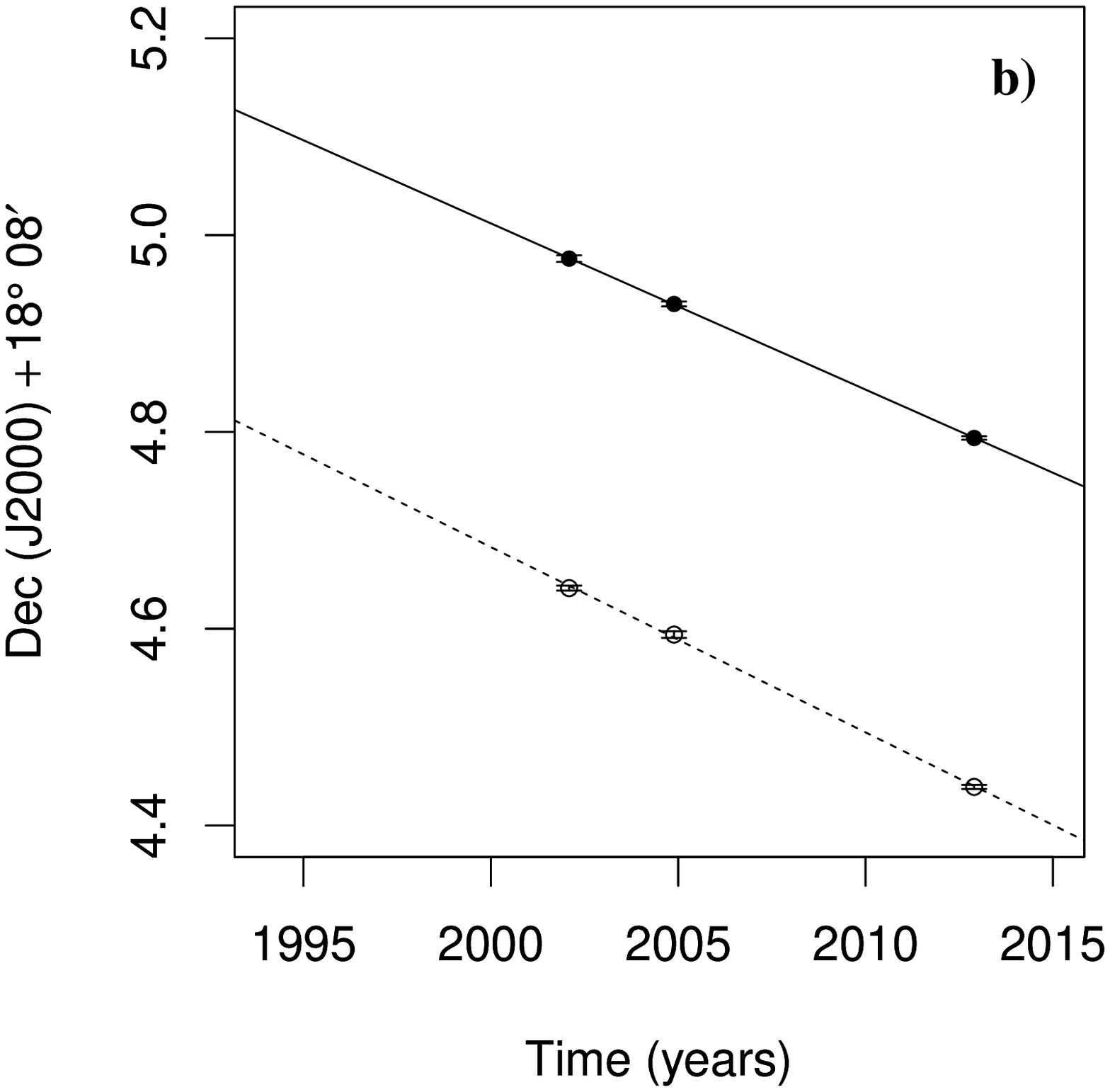}
   \caption{Same as Figure \ref{Set-1} for the subgroup 2 (data sets 3, 5 and 6).    }
  \label{Set-2}
\end{figure*}

In order to analyze each subset of observations that were carried out using the 
same calibrator, we also performed a linear least-squares fit to the measured
absolute positions of both components of L1551~IRS~5 as function of time
for each subset. The absolute proper motion of the L1551~IRS~5 system, using 
data sets with the phase calibrator 0431+206
(sets 1, 2 and 4), is on the order of 33.9~mas~yr$^{-1}$ (see Figure \ref{Set-1}), 
implying a velocity in the plane of the sky of $\sim 22.6$~km~s$^{-1}$.
On the other hand, an absolute proper motion of about 21.7~mas~yr$^{-1}$
($\sim 14.5$~km~s$^{-1}$; see Figure \ref{Set-2}) is estimated for the L1551~IRS~5 
binary system using the observations with the phase calibrator 0431+175
(sets 2, 5 and 6). 
Although the estimated proper motion for each subset is slightly different,
the mean of both values is very close to that estimated using all data sets.

\subsection{Relative Proper Motions}
\label{rpm}
\begin{figure}[ht]
    \includegraphics[angle=0,width=1.0\columnwidth]{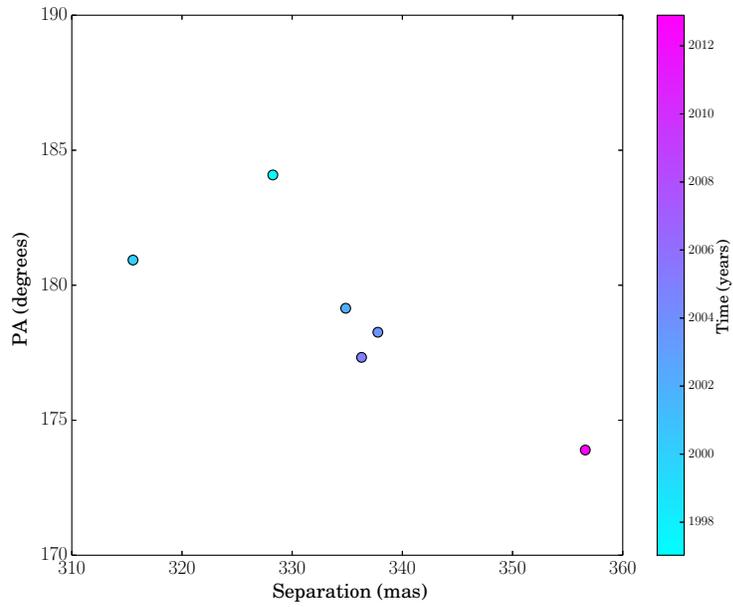}
 \caption{Plot (separation, position angle and time) of the temporal evolution
 of the projected separation of the component south with respect to component north
 and its position angle over a time span of 15 years
 of the  L1551~IRS~5 binary system.  }
  \label{Des-3D}
\end{figure}

In order to perform a detailed study of the relative proper motions of the L1551~IRS~5
 binary system, we have used all six data sets of 7~mm observations. First,
we have taken the spatial position of the north component (the most intense
component of the system) as reference and then we have measured the position
of the south component (right ascension and declination) and calculated its
projected separation for each observation epoch. After we have measured the
position angle between the north and south components of the binary system 
L1551~IRS~5. 
A three-dimensional plot (projected separation, position angle and time) is
shown in Figure \ref{Des-3D}, where relative proper motions of the two circumstellar
disks of the L1515~IRS~5 binary system are evident. This figure shows the change
over time of the spatial separation between the north and south components
as well as its position angle over a time span of about 15 years. A change in the 
projected separation between the two components is clearly observed, which is increasing
with time, while the position angle is decreasing.

\begin{figure*}[ht]
  \includegraphics[angle=0, width=0.55\linewidth,height=7.0cm]{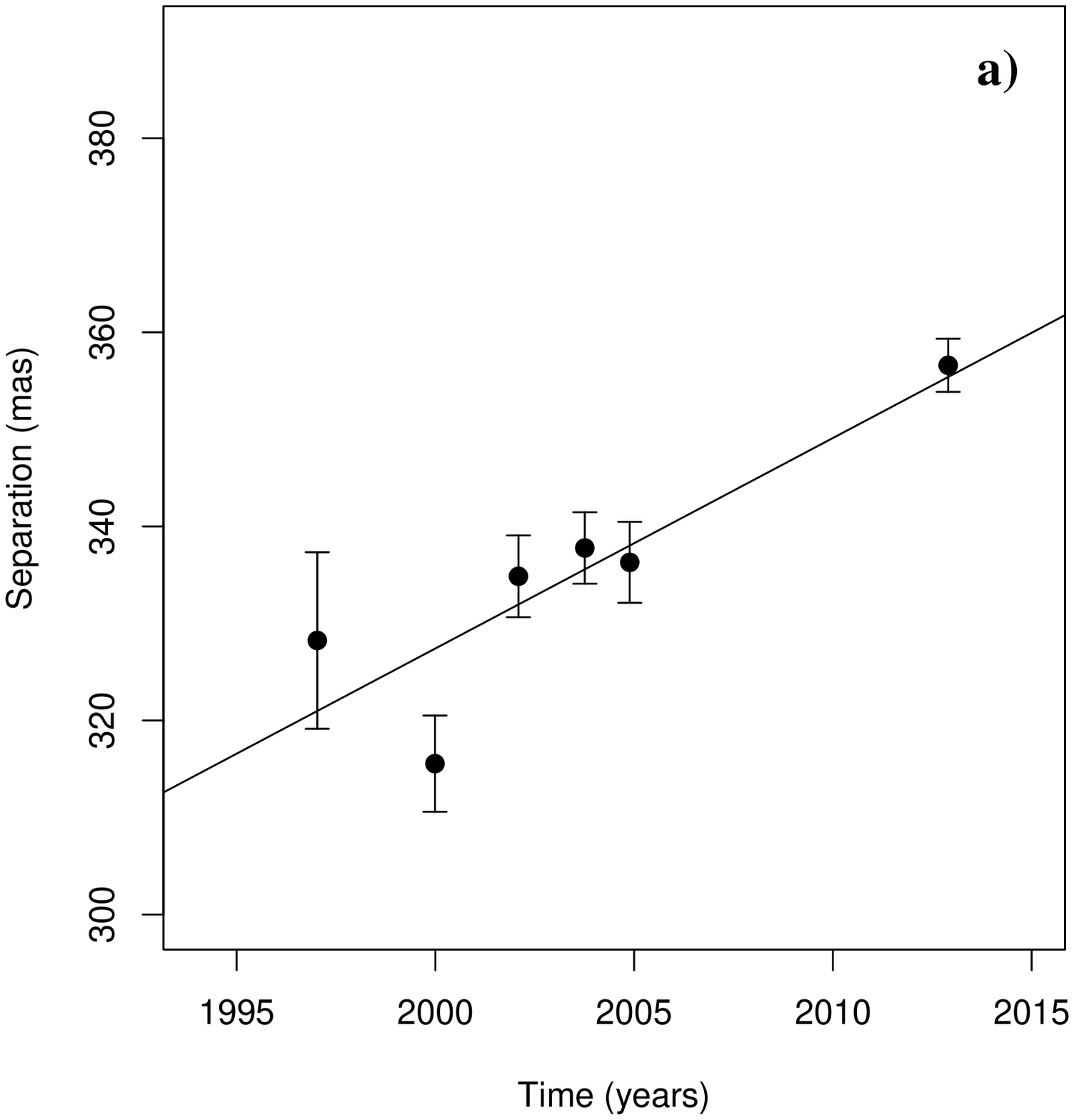}
 \hfill
  \includegraphics[angle=0, width=0.55\linewidth,height=7.0cm]{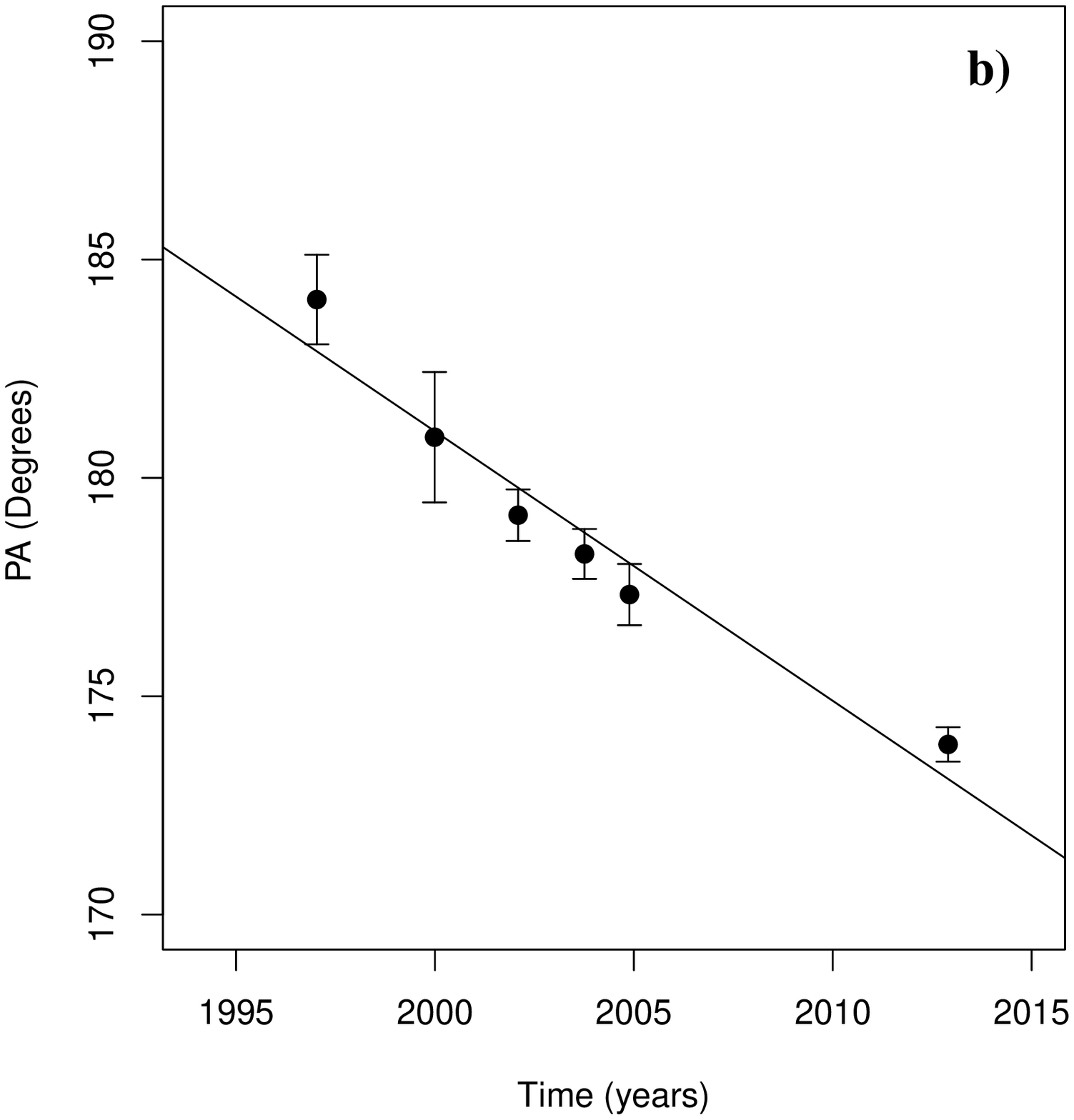}
   \caption{Temporal evolution of the projected separation of the compact circumstellar
   disks of L1551~IRS~5 (panel {\it a)}) and 
   position angle (panel {\it b)}).  Solid lines represent the least-squares fits.
   }
  \label{Sep-PA}
\end{figure*}

In order to calculate the relative proper motions, we made a linear least-squares
fit to the measured projected separation and the position angle between both components 
as function of time using all data sets. The results of these fits are shown
in Figure \ref{Sep-PA} (see also Table \ref{Table4}). From the fits  we found 
that the south component is being separated at a rate of $2.2\pm0.6$~mas~yr$^{-1}$,
while the position angle between the two components is changing to a rate of
$-0.6\pm0.1$~degree~yr$^{-1}$. On the other hand, using the six data sets of 7~mm 
observations and considering a time span of 15 years, we estimated an average
projected separation and position angle between the north and south components
of about $335\pm 5$~mas ($47 \pm 1$ assuming a distance of 140~pc)
 and $180 \pm 1^{o}$, respectively.
Using these values and those obtained from the fits, we calculated a relative proper
motion for the components  of the binary system L1551~IRS~5  on the order of 
4.2~mas~yr$^{-1}$, which is slightly larger than that estimated by \citet{Rodriguez2003},
who used 2~cm observations. However, as it is known, the circumstellar disk emission is
mainly produced by dust, and then the 7~mm observations trace a better way for this emission.
Therefore, we think that the relative proper motion of the compact circumstellar disks
of the L1551~IRS~5 binary system, as calculated using only 7~mm observations
is a value more accurate than that calculated using centimeter observations.

By the way, there is a point in the plot time versus separation falling away
from the fit in Figure \ref{Sep-PA}. This point corresponds to the observation 
carried out with the B configuration of the VLA. Since the observations in this 
VLA configuration has a lower angular resolution, the accuracy of the positions
will be lower; in addition to that these observations are sensitive to more 
extended structures, which could also affect the determination of the centroids
of the circumstellar disks.
These facts could indicates that only the 
observations with the A configuration could be the most appropriate for this type
of analysis. However, this single point does not affect substantially the result 
of the fit, since it only produces a very minimal change.

As it is known, the mass of a star is the most fundamental property, since it determines
how it will be its evolution and lifetime. However,
the determination of mass is one of the most difficult problems to solve in astronomy.
In particular, binary systems are one of the few cases where it is feasible to determine
the total mass of the system. Considering that the L1551~IRS~5 binary system is 
gravitationally bound and follows a Keplerian motion we can  estimate the 
total mass of the system, however it is necessary to make some considerations.
The orbital plane of the binary system and the plane of the disks are almost 
parallel \citep[see][]{Bate2000}, the orbit of the system is circular and the 
observed velocity is approximately $v_{real}/\cos i$, where $i$ is the inclination 
angle of the disks. Studies performed toward L1551~IRS~5 have estimated an
inclination angle of about 60$^{\circ}$ \citep[e.g.][]{Rodriguez1998, Chou2014}.
Under these considerations, we estimated a total mass of the binary system of
about $1.7\pm0.1$~M$_{\odot}$.
In addition to the total mass, we can also estimate the orbital
period of the components of the system using the Kepler's third law
, which is about $246\pm27$~years.
This estimated mass is consistent with that calculated for T Tauri binary systems
by \citet{Ghez1995} and \citet{Woitas2001}, who estimated masses
of about 1.7 and 2.0~M$_{\odot}$, respectively.

\section{Summary}
\label{summary}

In this paper we have analyzed 7~mm multi-epoch observations of the L1551~IRS~5 
binary system  using six data sets of the VLA archive. Observations were carried
out with the A (five sets) and B configurations covering a time span of about 15 years.
Given that the millimeter emission of L1551~IRS~5 comes mainly from the two compact
circumstellar disks, this wavelength represents a good way to calculate the proper
motions of the system.

We estimated an absolute proper motion of the L1551~IRS~5 binary system
on the order of 25.1~mas~yr$^{-1}$ using all data. This value should
be taken very carefully, since the observations were not performed with the
same phase calibrator, however, this value is consistent with those reported
by other authors. In addition, we found that the projected separation between
the two compact circumstellar disks is increasing with time and its position
angle is decreasing. Then, we calculated a relative proper motion of 
4.2~mas~yr$^{-1}$, which makes it possible to estimate total mass and orbital
period of the binary system, giving a value of 1.7~M$_{\odot}$ and 246 years,
respectively.

\acknowledgments
We would like to thank our referee for the very useful report on our
manuscript.
M.A.T. acknowledges support from DAIP (Universidad de Guanajuato).

\begin{table}[!ht]\centering
  \setlength{\tabnotewidth}{1.0\columnwidth}
  \tablecols{6}
  \caption{Observations from the VLA archive at 7~mm} 
  \label{Table1}
  \scriptsize
\begin{tabular}{ccccc}\toprule
Set &  Epoch    &    Project   &   Phase calibrator   &   Beam size        \\
    &           &          & (J2000)    &   (arsec)                \\ \midrule
1   &  1997.03  &  AR0277  &  0431+206  &   0.15$\times$0.09; -1.4      \\
2   &  1999.99  &  AT0235  &  0431+206  &   0.16$\times$0.12; -68.8      \\
3   &  2002.09  &  AT0269  &  0431+175  &   0.06$\times$0.04; -21.1      \\
4   &  2003.76  &  AC0675  &  0431+206  &   0.06$\times$0.05; -20.7      \\
5   &  2004.89  &  AC0743  &  0431+175  &   0.05$\times$0.05; 40.8      \\
6   &  2012.90  &  12B-091 &  0431+175  &   0.04$\times$0.04; -14.5      \\   \bottomrule

\end{tabular}
\end{table}

\begin{table}[!t]\centering
  \setlength{\tabnotewidth}{1.0\columnwidth}
  \tablecols{5}
  \caption{Position of the two components of L1551~IRS~5} 
  \label{Table2}
  \scriptsize
\begin{tabular}{ccccc}\toprule
Epoch   &   Component  &  R.A. (J2000)   &   Dec. (J2000)  &   \\  
        &              & 04$^{h}$ 31$^{m}$ & 18$^{\circ}$ 08$'$  &  \\  \midrule

1997.03  &  North  &  34.1411  &  05.075  &         \\
         &  South  &  34.1395  &  04.748  &         \\
1999.99  &  North  &  34.1422  &  04.997  &         \\
         &  South  &  34.1418  &  04.682  &         \\
2002.09  &  North  &  34.1494  &  04.976  &         \\
         &  South  &  34.1498  &  04.641  &         \\
2003.76  &  North  &  34.1527  &  04.946  &         \\
         &  South  &  34.1534  &  04.609  &         \\
2004.89  &  North  &  34.1515  &  04.930  &         \\
         &  South  &  34.1526  &  04.594  &         \\
2012.90  &  North  &  34.1576  &  04.794  &         \\
         &  South  &  34.1602  &  04.439  &         \\
         \bottomrule
\end{tabular}
\end{table}

\begin{table}[!t]\centering
  \setlength{\tabnotewidth}{1.0\columnwidth}
  \tablecols{4}
  \caption{Absolute proper motions of L1551~IRS~5} 
  \label{Table3}
  \scriptsize
\begin{tabular}{cccc}\toprule
 Data      &   Component  &  $\mu_{\alpha}$   &   $\mu_{\delta}$     \\
           &            &  (mas yr$^{-1}$) &  (mas yr$^{-1}$)   \\ \midrule
All data   &   North    &  +15.5$\pm$3.7   &   -17.1$\pm$0.8   \\
           &   South    &  +19.2$\pm$3.9   &   -19.2$\pm$0.4   \\       
Subgroup 1 &   North    &  +25.2$\pm$1.0   &   -18.8$\pm$2.0   \\
           &   South    &  +30.0$\pm$1.0   &   -20.6$\pm$2.0   \\
Subgroup 2 &   North    &  +10.8$\pm$1.0   &   -16.9$\pm$2.0   \\
           &   South    &  +13.7$\pm$1.0   &   -18.8$\pm$2.0   \\

         \bottomrule
\end{tabular}
\end{table}

\begin{table}[!t]\centering
  \setlength{\tabnotewidth}{1.0\columnwidth}
  \tablecols{5}
  \caption{Relative proper motions of L1551~IRS~5} 
  \label{Table4}
  \scriptsize
\begin{tabular}{ccccc}\toprule
Epoch    & $\Delta \alpha$ & $\Delta \delta$ & Separation & P.A.    \\
         &      (mas)      &      (mas)      &     (mas)  & (deg)    \\ \midrule
1997.03  & -23$\pm$6   & -327$\pm$9   &   328$\pm$9   &   184$\pm$1  \\
1999.99  &  -5$\pm$8   & -316$\pm$5   &   316$\pm$5   &   181$\pm$2  \\
2002.09  &   5$\pm$4   & -335$\pm$5   &   335$\pm$5   &   179$\pm$1  \\
2003.76  &  10$\pm$3   & -338$\pm$4   &   338$\pm$4   &   178$\pm$1  \\
2004.89  &  16$\pm$4   & -336$\pm$4   &   336$\pm$4   &   177$\pm$1  \\
2012.90  &  37$\pm$3   & -355$\pm$3   &   357$\pm$3   &   174$\pm$1  \\   \bottomrule

\end{tabular}
\end{table}

\end{document}